\documentstyle[12pt]{article}
\makeatletter
\newcommand{\singlespacing}{\let\CS=\@currsize\renewcommand{\baselinestretch}{1.0}\tiny\CS}
\newcommand{\doublespacing}{\let\CS=\@currsize\renewcommand{\baselinestretch}{1.5}\tiny\CS}

\begin{document}

\title{Conditionally exactly solvable potentials and 
supersymmetric transformations}
\author{G\'eza L\'evai$^{\rm a}$\thanks{levai@atomki.hu}, Pinaki Roy$^{\rm
b}$\thanks{pinaki@isical.ac.in}\\ 
$^{\rm a}$ Institute of Nuclear Research of the Hungarian\\
   Academy of Sciences, Debrecen, Pf. 51, Hungary 4001 \\
$^{\rm b}$ Physics and Applied Mathematics Unit\\ Indian Statistical 
   Institute\\ Calcutta, 700035, India}

\date{}

\maketitle

\begin{abstract}
A general procedure is presented to construct conditionally exactly 
solvable (CES) potentials using the techniques of supersymmetric 
quantum mechanics. The method is illustrated with potentials 
related to the harmonic oscillator problem. Besides recovering known 
results, new CES potentials are also obtained within the framework of 
this general approach. The conditions under which this method leads 
to CES potentials are also discussed. 
\end{abstract}

\section{Introduction}

Solvable potential problems have have played a dual role  
since the beginnings of quantum mechanics. First, they represented  
useful aids in modelling realistic physical problems, and second, 
they offered an interesting field of investigation in their own 
right. Related to this latter area, the concept of solvability 
has changed to some extent in recent years. Besides exactly 
solvable problems, for which the bound-state energy spectrum and 
solutions could be given in general analytical form, quasi-exactly 
solvable (QES) (see e.g. \cite{qes}) and conditionally exactly solvable 
(CES) \cite{ces} potential classes have also been identified recently.
   
In the first case only a finite number of eigenstates can be 
obtained exactly, while in the latter one analytical solutions are 
available only if some (or all) of the potential parameters are 
fine tuned to specific numerical values. 

There are different types of CES potentials, which is also reflected 
in the way they can be most naturally constructed. Some of them, including 
the first CES potentials \cite{ces,dutt}(see also ref \cite{znojil}
for some interesting comments on the construction of CES potentials)
have their bound-state solutions in terms of a single special function (polynomial); a structure 
characteristic of Natanzon class potentials \cite{natanzon}. For another 
class, these solutions have more complex structure, but generally 
they can be expressed in terms of the linear combination of two 
special functions. Typically these CES potentials are constructed as 
supersymmetric partners of some simple potentials \cite{gjpr,gjpr98}.   
Their CES nature hinges on the fact whether the parameters of their  
partners can be chosen such that they can be reduced to some 
simple potential with known solutions and energy eigenvalues. 
According to the techniques of supersymmetric quantum mechanics,  
or SUSYQM (for reviews see e.g. \cite{susyqm}), the CES potentials 
constructed in this way are then essentially isospectral with their 
partners, i.e. the two spectra are identical or differ only in their 
ground state. The bound-state solutions of CES potentials are obtained from 
those of their simple (Natanzon-type) partner potentials by acting on 
these latter ones with linear differential operators. This explains why 
their bound-state solutions possess the specific structure described above. 

In Refs. \cite{gjpr,gjpr98} some CES potentials have been constructed 
by SUSYQM. The aim of this publication is to show that this procedure 
can be made more systematic by making use of various types of 
SUSYQM transformations. The rather general nature of this treatment 
allows the recovery of known results and also the derivation of 
new CES potentials in the same framework. Our examples concern 
CES potentials related to the harmonic oscillator potential in three 
or one dimension (the standard examples of Refs. \cite{gjpr,gjpr98}), 
but the formalism is equally applicable to other types of potentials 
as well.

\section{The conventional SUSYQM approach to CES potentials}
\label{usual}

Let us assume that there is a pair of SUSYQM partner potentials 
$V_{\pm}^{(0)}(r)$, which can 
be constructed from a superpotential $W_0(r)$ in the usual way: 
\begin{equation}
V_{\pm}^{(0)}(r)=\frac{1}{2}[W_0^2(r)\pm W_0'(r)]\ .
\label{v0p}
\end{equation}
Consider now a superpotential of the form
\begin{equation}
W(r)=W_0(r)+w(r) \ . 
\label{wr}
\end{equation}
The partner potentials generated from $W(r)$ are then 
\begin{equation}
V_+(r)=V_+^{(0)}(r)+W_0(r)w(r)+\frac{1}{2}[w^2(r)+w'(r)]
\label{vp}
\end{equation}
\begin{equation}
V_-(r)=V_-^{(0)}(r)+W_0(r)w(r)+\frac{1}{2}[w^2(r)-w'(r)]\ .
\label{vm}
\end{equation}
Let us now insist on that one of these potentials, say $V_+(r)$ 
is related to some known potential up to an energy shift. In the 
simplest case this could be $V_+^{(0)}(r)$ in Eq. (\ref{v0p}):
\begin{equation}
V_+(r)=V_+^{(0)}(r)+\Delta\ .
\label{vpev0ppd}
\end{equation}
Combined with (\ref{vp}), this requirement immediately introduces a 
Riccati-type differential equation for $w(r)$:
\begin{equation}
\frac{1}{2}[w^2(r)+w'(r)]+W_0(r)w(r)=\Delta\ .
\label{wdiffeq}
\end{equation}
If this equation is solved, then a pair of SUSYQM potentials is obtained, 
from which one of the partner potentials, $V_+(r)$, corresponds to a known 
potential (up to an energy shift). Therefore, both the spectrum and the 
wavefunctions of the partner potential $V_-(r)$ can be obtained in 
the usual way. 

In the examples in \cite{gjpr} $V_+^{(0)}(r)$ was the harmonic oscillator 
potential in 1 and 3 dimensions, with $W_0(r)$ being the corresponding 
superpotential. In both cases the structure of $w(r)$ was of the type 
\begin{equation}
w(r)=\sum_{i=1}^N \frac{2g_i r}{1+g_i r^2}\ . 
\label{wjp}
\end{equation}
In the practical examples N=1 was used. The difference was that in the 
one-dimensional case the authors of \cite{gjpr} considered unbroken 
supersymmetry (and therefore $V_-(x)$ had one more state than $V_+(x)$), 
while in the three-dimensional case they chose to discuss broken 
supersymmetry, so the spectra of the partner potentials were identical.

\section{An alternative SUSYQM construction of CES potentials}
\label{alternat}

Here we use various SUSYQM transformations systematically to 
recapitulate the formalism of Section \ref{usual} and to put it 
into a more general context. 

It is known from the theory of isospectral potentials that a potential 
$V_2(r)$ isospectral with a known potential $V_1(r)$ can be constructed by 
\cite{sukumar}
\begin{equation}
V_2(r)=V_1(r)-\frac{{\rm d}^2}{{\rm d}r^2}\ln \phi(r)\ ,
\label{v2v1}
\end{equation}
where $\phi(r)$ is a solution of the Schr\"odinger equation 
\begin{equation}
-\frac{1}{2} \frac{{\rm d}^2 \phi}{{\rm d}r^2} + V_1(r)\phi =\epsilon \phi \ .
\label{schphi}
\end{equation}
Here $\epsilon$ is usually called factorization energy. Depending on the 
value of $\epsilon$ and the boundary conditions of the solution $\phi(r)$, 
$V_2(r)$ in (\ref{v2v1}) will have various properties. In the case of 
a radial problem (in three-dimensions) four types of 
transformations are possible. These are related four different types 
of {\it nodeless} solutions $\phi(r)$ of Eq. (\ref{schphi}) and 
have been described in terms of SUSYQM \cite{sukumar,baye,balaton}. 
The nodelessness of $\phi(r)$ guarantees that the resulting potential 
$V_2(r)$ does not have singularities for finite values of $r$ (besides 
the origin), and this can be achieved whenever the factorization 
energy $\epsilon$ is below the ground-state energy of $V_1(r)$ 
\cite{sukumar}. We briefly summarize the basic characteristics 
of the four SUSYQM transformation types in Table \ref{tab1}. 

Let us consider the radial harmonic oscillator as an example 
and solve (\ref{schphi}) for $\phi(r)$ with
\begin{eqnarray} 
V_1(r)=V_+^{(0)}(r)&=&\frac{1}{2}[W_0^2(r)+W_0'(r)]
\nonumber \\
&=&\frac{1}{2}r^2+ \frac{\gamma(\gamma+1)}{2r^2}+\gamma+\frac{3}{2}\ .
\label{v1r}
\end{eqnarray} 
Here the superpotential is $W_0(r)=r+(\gamma+1)r^{-1}$, and the bound 
states of $V_1(r)$ are found at $E_n=2n+2\gamma+3$.  
The solution $\phi(r)$ can be searched for in the form 
\begin{equation}
\phi(r)\simeq r^A\exp\left(\frac{B}{2}r^2\right)F(a,b;Cr^2)\ ,
\label{phir}
\end{equation}
where $F(a,b;z)$ is the confluent hypergeometric function \cite{as70}. 
Straightforward calculation shows that (\ref{schphi}) transforms into the 
confluent hypergeometric equation if the following conditions hold: 
\begin{eqnarray}
A(A-1)&=&\gamma(\gamma+1)\ , \hskip 1cm 
B^2=1\ , \hskip 1cm 
B=-C\ ; \label{cond1} \\
b&=&A+\frac{1}{2} \label{cond2} \\
a&=&-\frac{\epsilon}{2C}+\frac{\gamma}{2C}+\frac{3}{4C}
+\frac{A}{2}+\frac{1}{4}\ .
\label{cond}
\end{eqnarray}
Recalling that besides $F(a,b;z)$, $z^{1-b}F(a-b+1,2-b;z)$ is a 
linearly independent solution of the same confluent hypergeometric 
function \cite{as70}, the general solution $\phi(r)$ has the form 
\begin{eqnarray}
\phi(r)\simeq \exp\left(\frac{B}{2}r^2\right)[ 
&&\alpha_1 r^{\gamma+1}F(\frac{\epsilon}{2C}+\frac{\gamma}{2C}
  +\frac{3}{4C}+\frac{\gamma}{2}+\frac{3}{4},\gamma+\frac{3}{2};Cr^2)
  \nonumber \\
&&
  +\alpha_2 r^{-\gamma}F(\frac{\epsilon}{2C}+\frac{\gamma}{2C}
  +\frac{3}{4C}-\frac{\gamma}{2}+\frac{1}{4},-\gamma+\frac{1}{2};Cr^2)
]\ .
\label{gensol}
\end{eqnarray}
Note that the two terms in Eq. (\ref{gensol}) are connected by the 
$\gamma\leftrightarrow 1-\gamma$ transformation, therefore it is 
enough to consider one of the solutions ($A=\gamma+1$ or $A=-\gamma$) 
of $A(A-1)=\gamma(\gamma+1)$ in Eq. (\ref{cond1}). 
The solutions corresponding to the transformations $T_1$, $T_2$, $T_3$ and 
$T_4$ in Table \ref{tab1} can then be identified by imposing the appropriate 
boundary conditions on $\phi(r)$. 

Substituting the $\phi(r)$ function in Eq. (\ref{v2v1}) one obtains an
expression for $V_2(r)$ in terms of $\phi'(r)$ and $\phi''(r)$. With the 
use of (\ref{schphi}), $V_2(r)$ can be expressed in terms of 
$V_1(r)$, $\epsilon$ and $\phi'/\phi$. In this last expression 
the first-order derivatives of two confluent hypergeometric functions 
occur, each of which can be expressed in terms of another confluent 
hypergeometric function \cite{as70}. This means that $V_2(r)$ can 
be expressed in a somewhat complicated, but closed analytic form. 
A special situation occurs when $a=-N$ or $a-b+1=-M$ holds. 
In this case one of the confluent hypergeometric functions occuring 
in (\ref{gensol}) reduces to an $N$-th or $M$-th order (generalized 
Laguerre \cite{as70}) polynomial of the argument. According to 
(\ref{cond}), this case corresponds to specific choices of the 
factorization energy $\epsilon$. We note that in principle both confluent hypergeometric functions can reduce to a polynomial if both $\epsilon$ and 
$\gamma$ has a specific values: $\epsilon=-C(M+N+1)-\gamma-\frac{3}{2}$, 
$\gamma=M-N+\frac{1}{2}$. Let us now consider the four 
transformations $T_1$, $T_2$, $T_3$ and $T_4$ one by one.

\subsection{The $T_1$ case}
\label{t1}

The boundary condition at $r=0$ allows solutions only with $\alpha_2=0$ 
in (\ref{gensol}) (if $\gamma > 0$ holds), furthermore $B=-1$ is also 
required to fulfil the asymptotic boundary condition. The $a=-N$ 
condition leads to $\epsilon=2N+2\gamma+3$. This factorization energy corresponds to the bound-states energies of $V_1(r)=V_+^{(0)}(r)$ and 
$\phi(r)$ simply reproduces the physical wavefunctions. It is known that 
for $N=n\ne 0$ the transformed potential $V_2(r)$ has singularities at 
those locations, where the wavefunctions have nodes. The ground-state 
wavefunction with $n=0$, however, is nodeless, and the $T_1$ 
transformation then simply retrieves the classic SUSYQM transformation 
which eliminates the ground state of $V_1(r)$ and increases the value 
of $\gamma$ with one unit.

\subsection{The $T_3$ case}
\label{t3}

Similarly to the $T_1$ case only the term regular at the origin 
is allowed by the boundary condition at $r=0$ (i.e. $\alpha_2=0$), 
however, the asymptotic boundary condition requires $B=1$ in this case. 
The $a=-N$ polynomial condition then leads to the specific factorization 
energies $\epsilon=-2N$, which are always below the ground-state 
energy of $V_1(r)$, so the nodelessness of $\phi(r)$ is always 
secured. The $N=0$ choice recovers $V_2(r)$ 
as another oscillator with the same spectrum as $V_1(r)$: only the value 
of $\gamma$ is increased with one unit and the energy is shifted downwards 
with one unit. The $N=1$ case results in the CES potential described in 
\cite{gjpr} (denoted by $V_-(r)$ there) up to an energy shift:
\begin{equation}
V_2(r)=\frac{1}{2}r^2+\frac{(\gamma+1)(\gamma+2)}{2r^2}+\gamma+\frac{1}{2}
+\frac{4g_1^2r^2}{(1+g_1r^2)^2}-\frac{2g_1}{1+g_1r^2} \ ; \hskip 1cm 
g_1=\frac{2}{2\gamma+3} \ .
\label{v2t3n1}
\end{equation}
The energy shift is two units here, and it appears both in the numerical 
constant in (\ref{v2t3n1}) (it is $1/2$ here and $7/2$ in \cite{gjpr}) and 
the factorization energy ($-2$ here and $0$ in \cite{gjpr}). 
Similar, but more complicated isospectral potentials would arise from 
choosing $N>1$.

\subsection{The $T_4$ case}
\label{t4}

In contrast with the previous two cases, the boundary condition at the 
origin now allows both the regular and the singular solution in 
(\ref{gensol}). The ratio of the two coefficients, $\alpha_1$ and 
$\alpha_2$ appears as a new parameter in $V_2(r)$: the resulting 
potential family will have the same spectrum, but different shape. 
Similarly to the $T_3$ case, this one is usually also interpreted as 
a situation with broken supersymmetry, because the spectra of the 
partner potentials (and, of course, of the whole family) is identical. 

In order to get a situation similar to the $T_3$ and $T_1$ cases, 
we can restrict this potential family to a single potential, i.e. 
to that with $\alpha_1=0$ in (\ref{gensol}). The structure of 
$\phi(r)$ then becomes the same as before: it will contain only 
a single confluent hypergeometric function. With the loss of the 
generality, however, the nodelessness of $\phi(r)$ cannot be 
guaranteed in general, rather it has to be checked in each case 
separately. 

Taking also into account the asymptotic boundary condition 
which now requires $B=-1$, we find that the $a-b+1=-N$ polynomial 
condition now leads to factorization energies $\epsilon=2N+2$. 
The $N=0$ 
choice again results in another harmonic oscillator potential, with 
$\gamma$ decreased with one unit and with an energy shift of one unit 
upwards. 
For $N=1$ a potential similar to that in \cite{gjpr} arises, 
whenever $\gamma > 1/2$ holds. (As we have mentioned already, this 
latter condition secures that the polynomial 
$F(-1,-\gamma+\frac{1}{2};r^2)= 1+2r^2/(2\gamma-1)$
remains nodeless, and there will be no singularities in the $V_2(r)$. 
In fact, the $2\gamma+3=E_0 > \epsilon=2N+2$ condition also leads to 
$\gamma > 1/2$ for $N=1$.) The potential is then 
\begin{equation}
V_2(r)=\frac{1}{2}r^2+\frac{\gamma(\gamma-1)}{2r^2}+\gamma+\frac{5}{2}
+\frac{4g_1^2r^2}{(1+g_1r^2)^2}-\frac{2g_1}{1+g_1r^2} \ ; \hskip 1cm 
g_1=\frac{2}{2\gamma-1} \ .
\label{v2t4n1}
\end{equation}
The functional form of $V_2(r)$ is essentially the same as that of 
(\ref{v2t3n1}), only the value of $\gamma$, the numerical constant and 
$g_1$ is different. Similarly to the $T_3$ case, further potentials 
isospectral with a harmonic oscillator can be constructed by choosing 
$N > 1$, but the nodelessness of $\phi(r)$ has to be checked in each 
case.

\subsection{The $T_2$ case}
\label{t2}

The situation here is the same as in the $T_4$ case: both 
the regular and the singular solutions are allowed by the boundary 
condition at the origin. This means, that we again have a whole 
family of potentials $V_2(r)$, which have the same spectrum and 
differ only in their shape. As before, we again restrict our attention 
to a particular member of this family, i.e. to the potential 
obtained with $\alpha_1=0$. Furthermore, we consider the polynomial 
condition $a-b+1=-N$, which leads to $\epsilon=-2N+2\gamma+1$. 

For $N=0$, 
$V_2(r)=\frac{1}{2}r^2+\frac{\gamma(\gamma-1)}{2r^2}+\frac{1}{2}$, 
which corresponds to another harmonic oscillator potential with the 
$\gamma$ value decreased by one unit and also shifted lower by one energy 
unit. Clearly, this corresponds to the usual SUSYQM transformation which 
inserts a new state (at $E=2\gamma+1$) below the ground state of $V_1(r)$. 
For $N=1$ we find that 
$F(-1,-\gamma+\frac{1}{2};-r^2= 1-2r^2/(2\gamma-1)$, which has 
a node at a positive value of $r$, unless $\gamma< 1/2$ holds. 
As in the $T_4$ case, here we have to check the nodelessness of 
$\phi(r)$ in each case, because it cannot be automatically 
guaranteed after we restricted the general solution by selecting 
$\alpha_1=0$ in (\ref{gensol}). The functional form of $V_2(r)$ is 
the same as (\ref{v2t4n1}), but with $g_1=2/(1-2\gamma)$.

\subsection{The one-dimensional case}
\label{1d}

We note that similarly to the radial equation, the one-dimesional case 
can also be handled in the present framework. The difference arises from the 
different boundary conditions. The various transformation types 
corresponding to solutions of a Schr\"odinger equation with different 
boundary conditions has been described in \cite{jmsdb}. Here we only 
mention the example discussed in \cite{gjpr} for one dimension. In order 
to construct $V_2(r)$ with one more bound state than $V_1(r)$, the 
solution diverging in both directions has to be considered \cite{jmsdb}. 
In general, such a solution should be constructed from the two 
linearly independent solutions of the Schr\"odinger equation, similarly 
to the situation seen in Subsections \ref{t4} and \ref{t2}. However, 
taking only one of these, 
$\phi(x) = \exp(\frac{x^2}{2})F(-1,\frac{1}{2}; -x^2) = 
\exp(\frac{x^2}{2})(1+2x^2)$, 
we obtain the new state introduced for $V_2(x)$ 
as $\psi^{(2)}_0(x)\simeq 1/\phi(x)$. Note that $V_2(r)$ is 
symmetric with respect to the $x\leftrightarrow -x$ transformation. 
In the general case $V_2(r)$ would be asymmetric, similarly to the 
SUSYQM partner potentials of the one-dimensional harmonic oscillator
constructed in Ref. \cite{sukumar1}.

\section{Discussion}

The relation of the two procedures outlined in Sections \ref{usual} and 
\ref{alternat} can be interpreted in a simple way by noting that 
the partner potentials are linked by $V_+(r)-V_-(r)=W'(r)$ and 
$V_1(r)-V_2(r)=(\ln \phi(r))''$. From this 
\begin{equation}
W(r)=(\ln \phi(r))' + c 
\label{wlnphi}
\end{equation}
follows. Direct integration of (\ref{wr}) and (\ref{wjp}) with      
$W_0(r)=r+(\gamma+1)r^{-1}$, as in \cite{gjpr} and $c=0$, indeed, recovers
the general solution $\phi(r)$ specific to the $T_3$ case:
\begin{equation}
\phi(r)\simeq r^{\gamma+1}\exp\left(\frac{r^2}{2}\right)
\Pi_{i=0}^N (1+g_i r^2)\ .
\label{phinew}
\end{equation}
In addition to the notation of \cite{gjpr}, $g_0=0$ was also introduced 
for convenience. This function is also an $N$'th order polynomial, as 
expected from (\ref{phir}) for $a=-N$. In addition to the $T_3$ case, 
the situation should be the same for the other cases mentioned here, 
including also the one-dimensional case. We note that in the $T_2$, $T_4$ 
and in the one-dimensional cases discussed here, (\ref{phinew}) is 
not the most general form of the solution, rather it is a specific 
member of a family of solutions obtained as the linear combination 
of two independent solutions. 

In fact, all the $V_2(r)$ potentials derived from polynomial-type 
$\phi(r)$ solutions can be expressed by a common formula. Substituting 
$a=-N$ in (\ref{phir}) and combining it with (\ref{v2v1}) and 
(\ref{v1r}) one gets 
\begin{equation} 
V_2(r)= \frac{1}{2}r^2+ \frac{\gamma(\gamma+1)+2A}{2r^2}+\gamma
        +\frac{3}{2}-B 
        -\frac{{\rm d}^2}{{\rm d}r^2}\ln F(-N,A+\frac{1}{2};Cr^2)\ .
\label{v2gen}
\end{equation} 
The solutions relevant to the $T_1$, $T_2$, $T_3$ and $T_4$ cases 
can then be obtained by substituting $[A,B,C]=[\gamma+1,-1,1]$, 
$[-\gamma,1,-1]$, $[\gamma+1,1,-1]$ and $[-\gamma,-1,1]$, respectively. 
In the $N=0$ case the last term in (\ref{v2gen}) cancels and $V_2(r)$ 
contains only terms characteristic of the three-dimensional harmonic 
oscillator potential. For $N=1$, $F(-1,A+\frac{1}{2};Cr^2)=1+g_1r^2$, 
with $g_1=-2C/(2A+1)$, which gives rise to two new terms, formally 
identical ith the last two terms of (\ref{v2t3n1}). 

Another interesting formula can also be derived if recalling (\ref{vpev0ppd}) 
and (\ref{v1r}), i.e. $V_1(r)=V_+(r)-\Delta$, which also indicates 
$V_2(r)=V_-(r)-\Delta$. From this $V_1(r)+V_2(r)=W^2(r)-2\Delta$ directly 
follows. On the other hand, (\ref{v2v1}) and (\ref{schphi}) 
lead to $V_2(r)+V_1(r)=2\epsilon +(\phi'/\phi)^2$. Combinig these two 
formulas, we get  
\begin{equation}
[W(r)-(\ln \phi(r))'][W(r)+(\ln \phi(r))']=2(\epsilon+\Delta)\ .
\label{w2p2}
\end{equation}
This means, that $W(r)=\pm (\ln \phi(r))'$ always requires $\epsilon=-\Delta$ 
to hold. In this way $\Delta$, the constant appearing in the Riccati 
equation (\ref{wdiffeq}) in Section \ref{usual} is related to the 
factorization energy used in the formulas in Section \ref{alternat}. 
Note that the $W=-(\ln \phi)'$ choice is also acceptable here, and it 
simply corresponds to the $V_+\leftrightarrow V_+$ and 
$V_1 \leftrightarrow V_2$ replacements. 

We stress that although the examples we presented here are 
related to the harmonic oscillator in three and one dimensions, our 
treatment can be applied to other types of potentials as well. 
Besides the relatively simple shape-invariant potential \cite{gen} 
(see also \cite{lg89,susyqm} for a list of them), any member of the 
more general Natanzon potential class \cite{natanzon} can also serve 
as the $V_+^{(0)}(r)$ reference potential in (\ref{vpev0ppd}). 
However, for the first such applications some more thoroughly studied 
special Natanzon potentials could be the best candidates \cite{specnat}.

\section*{Acknowledgments}
This work has been supported by by the OTKA grant No. F20689 (Hungary).

\begin{table}[t]
\begin{center}
\caption{SUSYQM transformations belonging to different types 
of solutions $\phi(r)$. The notation of Refs. [9,10,12] 
has been 
adapted to the potentials discussed here. 
}
\label{tab1}
\begin{tabular}{lcccc}
Transformation & $T_1$ & $T_2$ & $T_3$ & $T_4$ \\
\hline
$\epsilon$ & $\epsilon=E_0$ & $\epsilon < E_0$ & $\epsilon < E_0$ 
 & $\epsilon < E_0$ \\
lim$_{r\rightarrow 0}\phi$ & $r^{\gamma+1}$ & $r^{-\gamma}$  
 & $r^{\gamma+1}$ & $r^{-\gamma}$ \\
lim$_{r\rightarrow \infty}\phi$ & convergent & divergent  & 
 divergent  & convergent \\
Spectrum  & deletes ground state & adds new ground state 
 & none & none \\
\hskip .3cm modification &  & ($0 < \gamma$ only) & 
 & ($0 < \gamma$ only) \\
Singularity  & $(\gamma+1)r^{-2}$ & $-\gamma r^{-2}$
 & $(\gamma+1)r^{-2}$ & $-\gamma r^{-2}$ \\ 
\hskip .3cm modification &&&&\\
\end{tabular}
\end{center}
\end{table}

\end{document}